\documentclass[12pt]{article}
\usepackage[usenames,dvipsnames]{color}
\usepackage{amsmath,amsthm,amsfonts,amssymb,multicol,amscd,amsbsy}
\usepackage{graphicx}
\usepackage{array}
\usepackage{enumerate}
\usepackage[nodayofweek]{datetime}
\usepackage[english]{babel}
\usepackage{mathtools}
\usepackage[toc,page]{appendix}
\usepackage{verbatim} 
\usepackage{amsthm} 
\usepackage{enumitem}
\usepackage{enumerate}
\usepackage{bbm} 
\usepackage[normalem]{ulem} 
\usepackage{bm}
\usepackage{hyperref}
\usepackage{authblk}

\newcommand{\be}{\begin}
\newcommand{\e}{\end}
\newcommand{\beq}{\begin{equation}}
\newcommand{\eeq}{\end{equation}}

\renewcommand{\l}{\left}
\renewcommand{\r}{\right}

\newcommand{\x}{\mathbf{x}}
\newcommand{\y}{\mathbf{y}}

\newcommand{\n}{\mathbf{n}}

\newcommand{\set}[1]{\mathbb{#1}}
\newcommand{\curly}[1]{\mathcal{#1}}

\newcommand{\setof}[2]{\left\{ #1\; : \;#2 \right\}}

\newcommand{\C}{\set{C}}

\newcommand{\Lam}{\Lambda}
\newcommand{\gam}{\gamma}

\theoremstyle{definition}

\numberwithin{equation}{section}

\theoremstyle{remark}

\def\dotuline{\bgroup
  \ifdim\ULdepth=\maxdimen  
   \settodepth\ULdepth{(j}\advance\ULdepth.4pt\fi
  \markoverwith{\begingroup
  \advance\ULdepth0.08ex
  \lower\ULdepth\hbox{\kern.15em .\kern.1em}%
  \endgroup}\ULon}

\def\dashuline{\bgroup
  \ifdim\ULdepth=\maxdimen  
   \settodepth\ULdepth{(j}\advance\ULdepth.4pt\fi
  \markoverwith{\kern.15em
  \vtop{\kern\ULdepth \hrule width .3em}%
  \kern.15em}\ULon}
\allowdisplaybreaks

\usepackage{tikz}

\usetikzlibrary{decorations.pathreplacing,decorations.markings}
\usepackage{xcolor}
\usepackage{xspace}

\tikzset{
  mid arrow/.style={postaction={decorate,decoration={
        markings,
        mark=at position .5 with {\arrow[#1]{stealth}}
      }}},
}

\newcommand{\hu}{\tikz{\draw[thick] (0,0)--(0.2,0.2);\draw[fill=black] (0,0) circle (1pt);\draw[fill=black] (0.2,0.2) circle (1pt);}}
\newcommand{\hd}{\tikz{\draw[thick] (0,0.2)--(0.2,0);\draw[fill=black] (0,0.2) circle (1pt);\draw[fill=black] (0.2,0) circle (1pt);}}
\newcommand{\hz}{\tikz{\draw[thick] (0,0)--(0.28,0);\draw[fill=black] (0,0) circle (1pt);\draw[fill=black] (0.28,0) circle (1pt);}}

\newcommand{\hout}{\tikz{\draw[thick] (0,0)--(0.14,0.14)--(0,0.28);\draw[fill=black] (0,0) circle (1pt);\draw[fill=black] (0.14,0.14) circle (1pt);\draw[fill=black] (0,0.28) circle (1pt);}}
\newcommand{\hin}{\tikz{\draw[thick] (0.14,0)--(0,0.14)--(0.14,0.28);\draw[fill=black] (0.14,0) circle (1pt);\draw[fill=black] (0,0.14) circle (1pt);\draw[fill=black] (0.14,0.28) circle (1pt);}}
\newcommand{\hmix}{\tikz{\draw[thick] (0,0)--(0.14,0.14)--(0.34,0.14);\draw[fill=black] (0,0) circle (1pt);\draw[fill=black] (0.14,0.14) circle (1pt);\draw[fill=black] (0.34,0.14) circle (1pt);}}

\newcommand{\hconn}{\tikz{\draw[thick] (0,0)--(0.14,0.14);
\draw[thick] (0.34,0.14)--(0.48,0);\draw[fill=black] (0,0) circle (1pt);\draw[fill=black] (0.14,0.14) circle (1pt);\draw[fill=black] (0.34,0.14) circle (1pt);\draw[fill=black] (0.48,0) circle (1pt);}}

\begin{document}
\title{The AKLT model on a hexagonal chain is gapped}

\date{March 31, 2019}

\author[1]{Marius Lemm\thanks{mlemm@math.harvard.edu}}
\author[2,3]{Anders Sandvik\thanks{sandvik@buphy.bu.edu}}
\author[2]{Sibin Yang\thanks{sibiny@bu.edu}}

\affil[1]{\textit{Department of Mathematics, Harvard University, 1 Oxford Street, Cambridge, MA 02138, USA}}
\affil[2]{\textit{Department of Physics, Boston University, 590 Commonwealth Avenue, Boston, MA 02215, USA}}
\affil[3]{\textit{Beijing National Laboratory for Condensed Matter Physics and Institute of Physics, Chinese Academy of Sciences, Beijing 100190, China}}

%
%

\maketitle

\begin{abstract}
In 1987, Affleck, Kennedy, Lieb, and Tasaki introduced the AKLT spin chain and proved that it has a spectral gap above the ground state. Their concurrent conjecture that the two-dimensional AKLT model on the hexagonal lattice is also gapped remains open. In this paper, we show that the AKLT Hamiltonian restricted to an arbitrarily long chain of hexagons is gapped. The argument is based on explicitly verifying a finite-size criterion which is tailor-made for the system at hand. We also discuss generalizations of the method to the full hexagonal lattice.
\end{abstract}

\section{Introduction}
\subsection{Motivation}
In 1987, Affleck, Kennedy, Lieb, and Tasaki (AKLT) introduced a novel quantum spin chain and rigorously proved that it exhibits a spectral gap above the ground state sector \cite{AKLT87,AKLT88}. Their work was motivated by a well-known conjecture of Haldane \cite{H83a,H83b} that any integer-spin Heisenberg antiferromagnet in one dimension is gapped. The AKLT spin chain shares central physical features with the Heisenberg antiferromagnet (isotropy and an antiferromagnetic local interaction), but it differs from it in a key aspect which makes deriving a spectral gap for it more feasible: it is frustration-free, meaning that its ground states minimize all local spin interactions simultaneously. This insight of AKLT subsequently led to the investigation of quantum spin systems with matrix product ground states \cite{FNW}. In recent years, these ideas have been further developed into tensor network states and have become a staple of modern condensed-matter physics.

AKLT also defined analogous models on arbitrary $k$-regular graphs with local spin $k/2$. These higher-dimensional models are still frustration-free with explicit ground states of valence-bond type. AKLT focused on the case of the hexagonal lattice $\mathbb H$. Writing $\Lam$ for a subset of $\mathbb H$, one takes as the Hilbert space $\bigotimes_{j\in \Lam}\C^{4}$ (so each site carries spin $3/2$). The Hamiltonian is
\beq\label{eq:Hdefn}
H^{AKLT}_\Lam:=\sum_{j\sim k} P_{j,k}^{(3)},
\eeq
where $P_{j,k}^{(3)}$ denotes the projection onto total spin $3$ for each pair of neighboring vertices $j\sim k$ in $\Lam$ (subject to appropriate boundary conditions). In this setting, AKLT proved that truncated spin-spin correlations decay for ground states with periodic boundary conditions. This result was further refined by Kennedy, Lieb, and Tasaki \cite{KLT88}, who proved uniqueness of the infinite-volume ground state, thereby establishing that the hexagonal AKLT model does not display N{\'e}el order and is thus in a different phase than the corresponding antiferromagnetic Heisenberg model. 


Based on the above findings, \emph{AKLT conjectured that the hexagonal model is gapped as well}. Their conjecture remains open, despite the continued popularity of AKLT-type models, including for applications to quantum computation \cite{M,WAR11,WHR14}. There has been some recent progress towards the AKLT conjecture in \cite{Aetal}, where a spectral gap was derived for decorated AKLT models at sufficiently high decoration number ($\geq 3$).\\ 

\begin{figure}[t]
\begin{center}
\begin{tikzpicture}[scale=0.9]
\draw
(0:1)--(60:1)--(120:1)--(180:1)--(-120:1)--(-60:1)--cycle;
\foreach \n in {0, 60,120,180}{
\draw[fill=black] (\n:1) circle (2.5pt);}
\draw[fill=red] (-60:1) circle (2.5pt);
\draw[fill=green] (-120:1) circle (2.5pt);
\begin{scope}[shift={(0,1.73)}]
\draw
(0:1)--(60:1)--(120:1)--(180:1)--(-120:1)--(-60:1)--cycle;
\foreach \n in {0, 60, 120, 180, -120, -60}{
\draw[fill=black] (\n:1) circle (2.5pt);}
\end{scope}
\begin{scope}[shift={(0,3.46)}]
\draw
(0:1)--(60:1)--(120:1)--(180:1)--(-120:1)--(-60:1)--cycle;
\foreach \n in {0, 60, 120, 180, -120, -60}{
\draw[fill=black] (\n:1) circle (2.5pt);}
\end{scope}
\begin{scope}[shift={(0,5.19)}]
\draw
(0:1)--(60:1)--(120:1)--(180:1)--(-120:1)--(-60:1)--cycle;
\foreach \n in {0, 180, -120, -60}{
\draw[fill=black] (\n:1) circle (2.5pt);};
\draw[fill=red] (60:1) circle (2.5pt);
\draw[fill=green] (120:1) circle (2.5pt);
\end{scope}
\end{tikzpicture}
\end{center}
\caption{The hexagonal chain $\mathbb H_n$ for $n=4$. The boundary conditions are periodic in the long direction, so that the two green vertices, and the two red vertices, are identified.}
\label{fig:chain}
\end{figure}
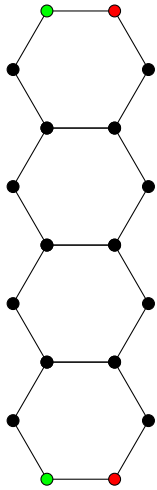

In this paper, we consider the original AKLT model, i.e., a spin-$3/2$ antiferromagnet, restricted to a chain of hexagons as in Figure \ref{fig:chain}. Our main result is that this Hamiltonian is gapped. More precisely, we prove that there is an explicit numerical constant that bounds the gap of the hexagonal chain from below, independently of the length of the chain. 

 Our proof is based on verifying a tailor-made finite-size criterion in the spirit of Knabe \cite{K} by an explicit computer calculation. Recently, finite-size criteria have been further developed in related contexts \cite{GM,L1,LM} and here we show that this technique does say something (but not everything one wants) about the hexagonal AKLT model that had originally motivated Knabe. The present work is the first instance where such a finite-size criterion can be verified by an exact computation of the eigenvalues of a subsystem with genuine 2D features. At the end of the paper, we describe how one can extend the method to the full hexagonal lattice. While we have not been able to verify the relevant finite-size criteria, we identify a certain class of subsystems of the hexagonal lattice which we believe are the most promising ones for establishing the AKLT conjecture by this method.

\subsection{Model and main result}
We write $\mathbb{H}_n$ for a vertical chain comprised of $n$ hexagons as in Figure \ref{fig:chain}. On the Hilbert space $\curly{H}_n=\bigotimes_{j\in \mathbb H_n}\C^{4}$, we consider the AKLT Hamiltonian \eqref{eq:Hdefn}. More specifically, we define
\beq\label{eq:Hndefn}
H_n=\sum_{\substack{j,k\in \mathbb H_n \\ j \leftrightarrow k}} P_{j,k}^{(3)}
\eeq
Here the neighbor relation $j\leftrightarrow k$ between sites $j$ and $k$ is defined such that $\mathbb H_n $ has open boundary conditions in the horizontal direction, and periodic boundary conditions in the vertical direction; see the endpoints in Figure \ref{fig:chain}. 

Note that $H_n\geq 0$ as an operator. From the work of AKLT \cite{AKLT87,AKLT88}, we know that $H_n$ is frustration-free, which in this normalization just means that its ground state energy is zero: $\ker H_n\neq \{0\}$. The spectral gap above the ground state is defined to be the smallest positive eigenvalue of $H_n$, and we denote it by $\gam_n>0$.

Our main result says that $H_n$ is gapped in the following strong sense.

\be{thm}[Main result]
\label{thm:main}
There exists a universal constant $c>0$ such that for every $n\geq 2$, 
$$
\gam_n\geq c>0.
$$
\e{thm}

We remark that for all $n\geq 30$, one can take the explicit constant $c=0.015$.

\subsection{Discussion of methods}
The proof of Theorem \ref{thm:main} is based on numerically verifying a tailor-made finite-size criterion (Theorem \ref{thm:fs}). The main novelty of the finite-size criterion we use here is that it involves three different types of subsystems, and some of these are included with multiplicity when constructing the full Hamiltonian $H_n$. 

We recall from \cite{GM,K,LM} that the choice of the appropriate subsystems to use for the finite-size criterion in dimensions $\geq 2$ is a delicate matter, since it has to negotiate several competing requirements: 
\begin{itemize}
\item[(a)] the subsystems have to collectively contain all different pair of neighboring edges the same number of times, 
\item[(b)] they should be large enough to allow for an efficient covering of the whole system,
\item[(c)] they should be small enough to be accessible to computer calculations. 
\e{itemize}
We remark also that the counting mentioned in point (a) does not in fact need to match perfectly, since a potential offset can be removed by the operator Cauchy-Schwarz inequality as in \cite{L1,LN}. Nonetheless, this comes at the price of making the covering achieved less efficient, and is thus inadvisable for very small subsystems in view of (b). 


A key difference of the present work compared to the recent work \cite{Aetal} is that there is no free parameter within the original AKLT model, so there is little flexibility in dealing with the explicit eigenvalues the original AKLT model presents us with here. The flexibility that we have is designing the subsystems which we use for the finite-size criterion. In \cite{Aetal}, there is a parameter in the model (called the ``decoration number'') that can be utilized effectively to prove a gap for the full two-dimensional system.\\  

The paper is organized as follows. In Section \ref{sect:1}, we prove the finite-size criterion, Theorem \ref{thm:fs} below. In Section \ref{sect:2}, we verify it by numerically calculating the spectral gaps of the relevant subsystems. In Section \ref{sect:3}, we describe our (unsuccessful) efforts towards deriving a spectral gap for the full hexagonal lattice (which would establish the original AKLT conjecture) by these methods. 

The main take-away in Section \ref{sect:3} is that we can identify certain small subsystems for which the finite-size criterion fails only narrowly. Now, these subsystems have bigger cousins which are currently not attainable by the precise computer calculations that we use here (Lanczos method and full diagonalization). Still, we can identify these bigger versions as promising candidates for verifying a finite-size criterion in the future. For instance, these systems may be amenable to DMRG (density matrix renormalization group) calculations as we aim to explore in a future work. Some previous results in this direction were obtained in \cite{GMW}.

\section{The finite-size criterion}
\label{sect:1}

\subsection{The subsystem Hamiltonians}
Let $n\geq 20$ from now on. We introduce the three basic kinds of finite subsystem that will feature in our finite-size criterion. We call these the $A$, $B$, and $C$ systems. Systems $A$ and $B$ are shown in Figure \ref{fig:subs}. Note that system C is simply a one-dimensional open chain on $K$ sites. To make sense of the $C$-subsystem for large values of $K$ (we will eventually take $K=14$), we assume that $n>2K$.

\begin{figure}[t]
\begin{center}
\begin{tikzpicture}
\node at (0, -.5)    {A};
\draw (0:1)--(60:1);
\draw (120:1)--(180:1);
\foreach \n in {0, 60,120,180}{
\draw[fill=black] (\n:1) circle (2.5pt);}
\begin{scope}[shift={(0,1.73)}]
\draw
(0:1)--(60:1)--(120:1)--(180:1)--(-120:1)--(-60:1)--cycle;
\foreach \n in {0, 60, 120, 180, -120, -60}{
\draw[fill=black] (\n:1) circle (2.5pt);}
\end{scope}
\begin{scope}[shift={(0,3.46)}]
\draw
(0:1)--(60:1)--(120:1)--(180:1)--(-120:1)--(-60:1)--cycle;
\foreach \n in {0, 60, 120, 180, -120, -60}{
\draw[fill=black] (\n:1) circle (2.5pt);}
\end{scope}
\begin{scope}[shift={(0,5.19)}]
\draw (-60:1)--(0:1);
\draw (180:1)--(-120:1);
\foreach \n in {0,  180, -120, -60}{
\draw[fill=black] (\n:1) circle (2.5pt);}
\end{scope}
\begin{scope}[shift={(5,+.8)}]
\node at (0, -1.3)    {B};
\draw
(0:1)--(60:1)--(120:1)--(180:1)--(-120:1)--(-60:1)--cycle;
\foreach \n in {0, 60,120,180,-120,-60}{
\draw[fill=black] (\n:1) circle (2.5pt);}
\begin{scope}[shift={(0,1.73)}]
\draw
(0:1)--(60:1)--(120:1)--(180:1)--(-120:1)--(-60:1)--cycle;
\foreach \n in {0, 60, 120, 180, -120, -60}{
\draw[fill=black] (\n:1) circle (2.5pt);}
\end{scope}
\begin{scope}[shift={(0,3.46)}]
\draw
(0:1)--(60:1)--(120:1)--(180:1)--(-120:1)--(-60:1)--cycle;
\foreach \n in {0, 60, 120, 180, -120, -60}{
\draw[fill=black] (\n:1) circle (2.5pt);}
\end{scope}
\end{scope}
\end{tikzpicture}
\end{center}
\caption{The $A$- and $B$-subsystems of $\mathbb H_n$.}
\label{fig:subs}
\end{figure}
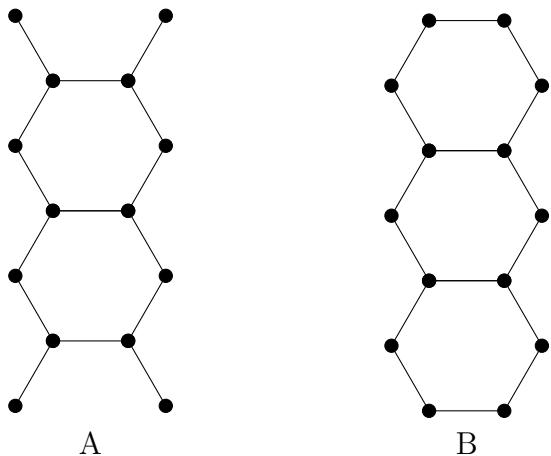

 To define the subsystems formally, it is convenient to represent the hexagonal chain using Cartesian basis vectors $(1,0)$ and $(0,1)$, so that $(0,1)$ points in what we called the vertical direction and the origin $(0,0)$ lies at a bottom left corner of one of the hexagons. See Figure \ref{fig:Cartesian}. Then we take as the vertex sets
$$
A:=\bigcup_{x=1}^7 \{ (0,x),(1,x)\},\qquad
B:=\bigcup_{x=0}^6 \{ (0,x),(1,x)\},\qquad
C:=\bigcup_{x=1}^{K} \{(0,x)\}.
$$
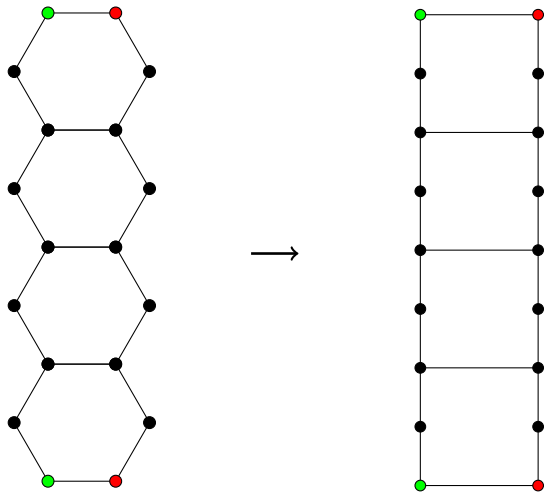
\begin{figure}[t]
\begin{center}
\begin{tikzpicture}[scale=0.9]
\draw
(0:1)--(60:1)--(120:1)--(180:1)--(-120:1)--(-60:1)--cycle;
\foreach \n in {0, 60,120,180}{
\draw[fill=black] (\n:1) circle (2.5pt);}
\draw[fill=red] (-60:1) circle (2.5pt);
\draw[fill=green] (-120:1) circle (2.5pt);
\begin{scope}[shift={(0,1.73)}]
\draw
(0:1)--(60:1)--(120:1)--(180:1)--(-120:1)--(-60:1)--cycle;
\foreach \n in {0, 60, 120, 180, -120, -60}{
\draw[fill=black] (\n:1) circle (2.5pt);}
\end{scope}
\begin{scope}[shift={(0,3.46)}]
\draw
(0:1)--(60:1)--(120:1)--(180:1)--(-120:1)--(-60:1)--cycle;
\foreach \n in {0, 60, 120, 180, -120, -60}{
\draw[fill=black] (\n:1) circle (2.5pt);}
\end{scope}
\draw [->,line width=0.35mm] (2.5,2.5)--(3.2,2.5);
\begin{scope}[shift={(0,5.19)}]
\draw
(0:1)--(60:1)--(120:1)--(180:1)--(-120:1)--(-60:1)--cycle;
\foreach \n in {0, 180, -120, -60}{
\draw[fill=black] (\n:1) circle (2.5pt);};
\draw[fill=red] (60:1) circle (2.5pt);
\draw[fill=green] (120:1) circle (2.5pt);
\end{scope}
\begin{scope}[shift={(5,-1.8)},scale=0.87]
\foreach \x in {0,2}{
\draw (\x,1)--(\x,9);
}
\foreach \y in {1,3,5,7,9}{
\draw (0,\y)--(2,\y);
}
\foreach \y in {2,3,4,5,6,7,8}{
\draw[fill=black] (0,\y) circle (2.58pt);
\draw[fill=black] (2,\y) circle (2.58pt);
}
\draw[fill=green] (0,1) circle (2.58pt);
\draw[fill=red] (2,1) circle (2.58pt);
\draw[fill=green] (0,9) circle (2.58pt);
\draw[fill=red] (2,9) circle (2.58pt);
\end{scope}
\end{tikzpicture}
\end{center}
\caption{The right picture is the Cartesian representation of the hexagonal chain $\mathbb H_4$ on the left side. Our convention is to place the origin $(0,0)$ of the Cartesian coordinate system at the bottom left corner.}
\label{fig:Cartesian}
\end{figure}

We write $\curly{E}_{\mathbb H_n}$ for the set of all edges in $\mathbb H_n$, which includes the edges coming from the periodic boundary conditions in the vertical direction. For any $\#\in \{A,B,C\}$ we write $\curly{E}_\#$ for the set of edges in $\#$, i.e., the set of unordered pairs $(j,k)$ with $j,k\in\#$ such that $j$ and $k$ are connected when $\#$ is viewed as a subgraph of $\mathbb H_n$. In other words, $\curly{E}_\#$ contains precisely the edges of $\mathbb H_n$ appearing in Figure \ref{fig:subs}.

From now on, we write
$$
h_e:=P_{e}^{(3)}
$$
for every edge $e\in \curly{E}_{\mathbb H_N}$. We define the subsystem Hamiltonians by
\beq\label{eq:Hsharpdefn}
H^{\#}:=\sum_{e\in \curly{E}_\#} h_e,\qquad \textnormal{ for any } \#\in\{A,B,C\}.
\eeq

The subsystem Hamiltonians $H^\#$ with $\#\in\{A,B,C\}$ are also frustration-free and we write $\gam^\#$ for their respective spectral gaps. We introduce the shorthand notation
$$
\gam_{\min}=\gam_{\min}(K):=\min_{\# \in\{A,B,C\}} \gam^\#
$$
(The $K$-dependence arises because $C$ is a chain on $K$ sites.)

Our proof rests on the following technical result.

\be{thm}[Finite-size criterion]\label{thm:fs}
Let $K\geq 4$ be an even integer, and let $n\geq \max\{20,2K+1\}$. Then
$$
\gam_{n}\geq \frac{7}{6} \l(\gam_{\min}-\frac{1}{7}-\frac{1}{7(K-2)}\r).
$$
\e{thm}

The criterion says that if for some fixed $K$, the gap of the subsystems $\gam_{\min}=\gam_{\min}(K)$ exceeds the threshold $\frac{1}{7}+\frac{1}{7(K-2)}$, then the spectral gap of the whole Hamiltonian $\gam_n$ has an $n$-independent positive lower bound.

\subsection{Squaring the Hamiltonian}
For the proof, we will denote $H_n=H$. The basic mechanism that drives the proof is that, thanks to frustration-freeness and the spectral theorem, the claimed bound on $\gam_n$ is equivalent to the operator inequality
\beq\label{eq:H2claim}
H^2\geq \frac{7}{6} \l(\gam_{\min}-\frac{1}{7}-\frac{1}{7(K-2)}\r) H.
\eeq

The goal is thus to prove \eqref{eq:H2claim}. We start by computing the left-hand side. Given two edges $e$ and $e'$, we write $e\sim e'$ if $e$ and $e'$ are distinct, but share a vertex, and we write $e\not\sim e'$, if $e$ and $e'$ are distinct and do not share a vertex. We also introduce the anticommutator of two operators $A$ and $B$,
$$
\{A,B\}=AB+BA.
$$
Using that $h_e^2=h_e$, we obtain
\beq\label{eq:H2}
H^2=H+Q+R
\eeq
with
\beq\label{eq:QRdefn}
Q:=\sum_{\substack{e,e'\in \curly{E}_{\mathbb H_n}:\\ e\sim e'}} \{h_e,h_{e'}\},
\qquad R:=\sum_{\substack{e,e'\in \curly{E}_{\mathbb H_n}:\\ e\not \sim e'}} \{h_e,h_{e'}\}.
\eeq

\subsection{Shifted subsystems and the auxiliary operator}
We introduce the reflected partner to subsystem $C$ and its associated Hamiltonian:
$$
\tilde C:=\setof{(1,x)\in \mathbb H_n}{(0,x)\in C}, \quad
\textnormal{ and }\quad
H^{\tilde C}:=\sum_{e\in \curly{E}_{\tilde C}} h_e.
$$
Here we used the Cartesian coordinate systems from Figure \ref{fig:Cartesian}.

Let us now fix two integers $n$ and $K$ with $n>\max\{20,2K\}$ and $K$ even. We translate the subsystem Hamiltonians $H^\#$ across the hexagonal chain, using the Cartesian representation throughout. Let $s$ be an integer. The subsystem $\#$ shifted upwards by $s$ units is given by
$$
\#+s:=\setof{(x,y)\in \mathbb H_n}{(x,y-s)\in \#}, \qquad \#\in\{A,B,C,\tilde C\}.
$$
For example, $A+1$ is the set $A$ shifted up by one unit in the vertical direction. 

We now define the associated shifted subsystem Hamiltonians. Recall that per the above definition, the edge set $\curly{E}_\#$ is a subset of $\mathbb H_n$, so the edges wrap around after they reach the $2n$th site (at which point periodic boundary conditions are enforced along the vertical boundary of $\mathbb H_n)$. We define the edge set $\curly{E}_{\#+s}$ analogously, i.e., it consists of all edges in the shifted system $\#+s$ subject to periodic boundary conditions at the top and bottom of $\mathbb H_n$.

The associated shifted subsystems Hamiltonians are given by
$$
H^{\#}_s=\sum_{e \in \curly{E}_{\#+s}} h_e,\qquad \textnormal{ for all } \#\in\{A,B,C,\tilde C\}.
$$
Note that all $H^{\#}_s$ are frustration-free, and by translation-invariance their spectral gaps are equal to the spectral gap $\gam^\#$ of the respective non-shifted Hamiltonian $H^\#$. Moreover, by reflection symmetry, $\gam^{\tilde C}=\gam^{C}$. 

We define the following auxiliary operator $\curly{A}$ in which we translate the subsystem Hamiltonians all over the hexagonal chain $\curly{H}_n$ (which we note has $2n$ sites):
$$
\curly{A}:=\sum_{s=0}^{2n-1} 
\l(  
(K-2)((H^A_{s})^2+(H^B_{s})^2)+(H^C_{s})^2+(H^{\tilde C}_{s})^2.
\r)
$$

The following proposition is at the heart of the proof of Theorem \ref{thm:fs}.

\be{prop}\label{prop:Akey}
We have the two operator inequalities
\begin{align}
\label{eq:A1}
\curly{A}&\geq 	7(K-2)\gam_{\min}H,	\\
\label{eq:A2}
\curly{A}&\leq (7(K-2)+1) H+6(K-2) (Q+R).
\end{align}
\e{prop}

\textit{Proof of \eqref{eq:A1}.} Since the subsystem Hamiltonians $H_s^\#$ are frustration-free, the spectral theorem implies the operator inequality
$$
(H_s^\#)^2\geq \gam^\# H_s^\#,\qquad \textnormal{ for all } \#\in\{A,B,C,\tilde C\}.
$$
Here we also used the aforementioned translation symmetry of the spectral gaps. By the definition of $\curly{A}$ and positivity of all $H^\#$, we see that
\beq\label{eq:ALB}
\curly{A}\geq \gam_{\min}  \sum_{s=0}^{2n-1} 
\l(  
(K-2)(H^A_{s}+H^B_{s})+H^C_{s}+H^{\tilde C}_{s}\r).
\eeq
This step also uses that $\gam^C=\gam^{\tilde C}$. 

Note that the right-hand side in \eqref{eq:ALB} is a sum of local interaction terms $h_e$ with $e\in \curly{E}_{\mathbb H_n}$ and it remains to count how often each edge $e$ is represented in it. By translation and reflection symmetry, there are only three types of edges, which we denote by $\hu$, $\hd$, and $\hz$, respectively. We decompose the total Hamiltonian $H$ in this way
\beq\label{eq:Hdecompose}
H=H_{\hu}+H_{\hd}+H_{\hz},\qquad
 \textnormal{ with } H_{\hu}:=\sum_{\substack{e\in \curly{E}_{\mathbb H_n}:\\ e \textnormal{ is } \hu}} h_e
\eeq
and with analogous definitions of $H_{\hd}$ and $H_{\hz}$.

We first count the number of times that each $\hu$ edge appears on the right-hand side of \eqref{eq:ALB}: Each $\hu$ edge appears in $3$ shifted $A$-systems, in $3$ shifted $B$-systems, and in $K-1$ shifted $C$-system (or in $K-1$ shifted $\tilde C$-systems, if it lies on the right side of the chain). Altogether, including the factor $K-2$ in front of $H^A_{s}+H^B_{s}$, this yields the total count of $7(K-2)+1$. In other words, the right-hand side of \eqref{eq:ALB} contains the term $(7(K-2)+1)H_{\hu}$. We find the same count for each $\hd$ edge.

Next we count the number of times that each $\hz$ edge appears on the right-hand side of \eqref{eq:ALB}: Each $\hz$ edge appears in $3$ shifted $A$-systems, in $4$ shifted $B$-systems, and in $0$ of the $C$- and $\tilde C$-systems. This yields the total count $7(K-2)$. 

To summarize, these combinatorial considerations prove that
\beq\label{eq:ALBend}
\begin{aligned}
&\sum_{s=0}^{2n-1} 
\l(  
(K-2)(H^A_{s}+H^B_{s})+H^C_{s}+H^{\tilde C}_{s}\r)\\
&=(7(K-2)+1)(H_{\hu}+H_{\hd})+7(K-2)H_{\hz}\\
&=7(K-2)H+H_{\hu}+H_{\hd}
\end{aligned}
\eeq
Since $H_{\hu}+H_{\hd}\geq 0$, this can be combined with \eqref{eq:ALB} to prove the claim \eqref{eq:A1}.\\

\textit{Proof of \eqref{eq:A2}.} Using that $h_e^2=h_e$, we compute
$$
(H^\#_s)^2=H^\#_s+Q^\#_s+R^\#_s
$$
where $Q^\#_s$ and $R^\#_s$ are defined as in \eqref{eq:QRdefn}, but with the edges $e,e'$ now taken from the set $\curly{E}_{\#+s}$ instead of $\curly{E}_{\mathbb H_n}$. This allows us to decompose the operator $\curly{A}$ as follows:
\beq\label{eq:Iterms}
\begin{aligned}
\curly{A}
=&\sum_{s=0}^{2n-1} \l((K-2)(H^A_{s}+H^B_{s}))+H^C_{s}+H^{\tilde C}_{s}\r)\\
&+\sum_{s=0}^{2n-1} \l((K-2)(Q^A_{s}+Q^B_{s}))+Q^C_{s}+Q^{\tilde C}_{s}\r)\\
&+\sum_{s=0}^{2n-1} \l((K-2)(R^A_{s}+R^B_{s}))+R^C_{s}+R^{\tilde C}_{s}\r)\\
=&(I)+(II)+(III).
\end{aligned}
\eeq
The first sum was calculated in \eqref{eq:ALBend}. Using that $0\leq H_{\hz}$, we obtain
\beq\label{eq:combine1}
(I)\leq (7(K-2)+1) H.
\eeq

It thus remains to control the terms (II) and (III) in \eqref{eq:Iterms}. We summarize the result in the following lemma.

\be{lm}\label{lm:remains}
We have
\begin{align}
\label{eq:lm1}
(II)=6(K-2)Q.\\
\label{eq:lm2}
(III)\leq 6(K-2) R.
\end{align}
\e{lm}

Note that applying \eqref{eq:combine1} and the estimates in Lemma \ref{lm:remains} to \eqref{eq:Iterms} yields \eqref{eq:A2}, and hence Proposition \ref{prop:Akey}. It thus remains to prove the lemma.

\be{proof}[Proof of Lemma \ref{lm:remains}]
We first prove \eqref{eq:lm1}. By definition, term (II) in \eqref{eq:Iterms} is a sum of terms $\{h_{e},h_{e'}\}$ with the edges $e\sim e'$ sharing a vertex. By translation and reflection symmetry, there are three types of edge pairs $e\sim e'$, which we denote by $\hout$, $\hin$, and $\hmix$, respectively. In analogy to \eqref{eq:Hdecompose}, we decompose the term $Q$ defined in  \eqref{eq:QRdefn} as follows:
$$
Q=Q_{\hout}+Q_{\hin}+Q_{\hmix}, 
\qquad  \textnormal{ with }Q_{\hout}:=\sum_{\substack{e,e'\in \curly{E}_{\mathbb H_n}:\\ e\sim e'\\ (e,e') \textnormal{ is } \hout}} \{h_{e},h_{e'}\},
$$
and with analogous definitions of $Q_{\hin}$ and $Q_{\hmix}$.

We first count the number of times that each $\hout$ edge pair appears in (II): It appears in $2$ shifted $A$-systems, in $3$ shifted $B$-systems, and in $K-2$ shifted $C$-systems (or in $K-2$ shifted $\tilde C$-systems if it lies on the right side of the chain).

Taking into account the prefactor $K-2$ in front of the $A$- and $B$-systems in the definition of (II), we conclude that (II) contains the term $6(K-2)Q_{\hout}$. We find the same total count for the $\hin$ edge pairs.

Next we count the number of times that each $\hmix$ edge pair appears in (II): It appears in $3$ shifted $A$-systems and in $3$ shifted $B$-systems and not otherwise. Recalling the prefactor $K-2$, this again yields the total count $6(K-2)$. 

Altogether, these combinatorial considerations prove that $$
(II)=6(K-2)Q_{\hout}+6(K-2)Q_{\hin}+6(K-2)Q_{\hmix}=6(K-2)Q,
$$
i.e., \eqref{eq:lm1}.\\

It remains to prove \eqref{eq:lm2}. First, we observe that all terms contributing to (III) are of the form $\{h_e,h_{e'}\}$ with $e\not\sim e'$. Since $h_e$ and $h_{e'}$ commute in this case, we have $\{h_e,h_{e'}\}\geq 0$ and so all the involved terms are non-negative. Recalling the definition \eqref{eq:QRdefn} of $R$, we see that it suffices to show that every individual pair of edges $e\not\sim e'$ appears in (III) at most $6(K-2)$ times. 

The total count for pairs of edges $e\not\sim e'$ in (III) is indeed at most $6(K-2)$. Modulo translation and reflection symmetry, this bound is achieved by edge pairs $e\sim e'$ of the form $\hconn$. An important point here is that these edge pairs do not appear in the $C$- and $\tilde C$-systems. By the above considerations, this shows that $(III)\leq 6(K-2)R$.

This establishes \eqref{eq:lm2}, hence Lemma \ref{lm:remains}, and finishes the proof of Proposition \ref{prop:Akey}.
\e{proof}

\subsection{Concluding the finite-size criterion} 
We will now collect the previously established results to derive Theorem \ref{thm:fs}. We begin by recalling the claim \eqref{eq:H2claim} and formula \eqref{eq:H2} which says $H^2=H+Q+R$. Combining the two operator inequalities in Proposition \ref{prop:Akey}, we can eliminate the auxiliary operator $\curly{A}$ and find
$$
 (7(K-2)+1) H+6(K-2) (Q+R)\geq  7(K-2)\gam_{\min}H,
$$
or equivalently,
$$
Q+R\geq \frac{7}{6} \l(\gam_{\min}-1-\frac{1}{7(K-2)}\r) H.
$$
Applying this operator inequality to $H^2=H+Q+R$, we find
$$
H^2\geq \frac{7}{6} \l(\gam_{\min}-\frac{1}{7}-\frac{1}{7(K-2)}\r) H.
$$
This proves \eqref{eq:H2claim}, and hence Theorem \ref{thm:fs} by frustration-freeness and the spectral theorem.
\qed

\section{Verifying the finite-size criterion}
\label{sect:2}
By Theorem \ref{thm:fs}, the main result follows if we can find an even integer $K\geq 4$ such that the finite-size gap $\gam_{\min}(K)=\min_{\# \in\{A,B,C\}} \gam^\#$ exceeds the threshold:
\beq\label{eq:req}
\gam_{\min}(K)>\frac{1}{7}+\frac{1}{7(K-2)}.
\eeq

We are indeed able to verify \eqref{eq:req} for $K=14$ using numerical computations of the spectral gap of systems $A$, $B$, and $C$. The numerical values for these spectral gaps were computed via the standard Lanczos algorithm \cite{Sandvik}. With this method, one can use various symmetries for block-diagonalizing $H$ to reduce the computational effort.  In the present application the subsystems of interest do not have translational symmetry and the only useful spatial symmetries are the reflections about the central $x$- or $y$-axis of the hexagonal chains. Use of both of these reflection symmetries splits $H$ into four blocks. However, the reduction of the largest basis size roughly only corresponds to reducing the number of spins $S=3/2$ by one, and, thus, the increase in the  accessible system size is only marginal. 

To keep the computer program simple, we therefore did not use the reflection symmetries and only implemented the very simple block-diagonalization in the conserved magnetization in the $z$-direction, studying blocks with fixed magnetization individually. We tested the Lanczos results against full diagonalization for systems A,B, and C, and also found perfect agreement with previously published numerics \cite{K}. The results are summarized in Tables \ref{table:AB} and \ref{table:C}.

\begin{table}[t]
\begin{center}
\begin{tabular}{| c | c |}
\hline
Spectral gap & Lower bound\\
\hline
$\gam^A$  & 0.168\\
$\gam^B$ & 0.175 \\
\hline
\end{tabular}
\end{center}
\caption{The spectral gaps of the subsystem Hamiltonians $H^A$ and $H^B$. The numbers were obtained by the standard Lanczos algorithm and verified by a full diagonalization, then rounded down in the last decimal place to ensure the applicability of Theorem \ref{thm:fs}.}
\label{table:AB}
\end{table}

\begin{table}[t]
\begin{center}
\begin{tabular}{| c | c |}
\hline
Spectral gap $\gam^C$ for length $K$ & Lower bound\\
\hline
$\gam^C(5)$ & 0.388 \\
$\gam^C(10)$ & 0.337 \\
$\gam^C(11)$ & 0.333 \\
$\gam^C(12)$ & 0.330 \\
$\gam^C(13)$ & 0.329 \\
$\gam^C(14)$ & 0.327 \\
\hline
\end{tabular}
\end{center}
\caption{The spectral gaps of the subsystem Hamiltonians $H^C$ for various values of $K$. The numbers are obtained in the same way as those in Table \ref{table:AB}. We only use the $K=14$ value in the text.}
\label{table:C}
\end{table}

\be{proof}[Proof of Theorem \ref{thm:main}]
Let $K=14$. On the one hand, from Tables \ref{table:AB} and \ref{table:C}, we see that
$$
\gam_{\min}(K)=\gam^A> 0.168
$$
On the other hand, we have $\frac{1}{7}+\frac{1}{7(K-2)}=\frac{13}{84}<0.155$. Hence, we can apply Theorem \ref{thm:fs} with $K=14$ to see that
\beq\label{eq:almostdone}
\min_{n\geq 29}\gam_n> 0.015
\eeq
By definition, we have $\gam_n>0$ for any fixed $n\geq 2$. Hence, \eqref{eq:almostdone} implies that there exists a universal constant $c>0$ such that
$$
\min_{n\geq 2}\gam_n\geq c>0.
$$
This proves Theorem \ref{thm:main}.
\e{proof}

\section{Outlook: The AKLT conjecture}
\label{sect:3}
Our main result, Theorem \ref{thm:main}, establishes that the AKLT Hamiltonian on an infinite hexagonal chain is gapped. While this model incorporates two-dimensional features, it is quasi-one-dimensional, and therefore still far from the AKLT model on the full hexagonal lattice which is the one conjectured to be gapped in \cite{AKLT87,AKLT88}.

We also attempted to use the method for the AKLT Hamiltonian on the full hexagonal lattice \eqref{eq:Hdefn}. However, we have found that for the relatively small subsystems whose spectral gaps can be computed, the relevant finite-size criteria simply do not hold. For one particular kind of subsystem, which we call the ``hexagonal sun'' (see Figure \ref{fig:sun}), a certain refined finite-size criterion only fails somewhat narrowly, as we describe now.\\

We use weighted subsystem Hamiltonians to derive a refined finite-size criterion. This idea is originally due to Kitaev and was developed in \cite{GM,LM}, but with a focus on large subsystems. Here we apply the weighting method for very small subsystems. (We thus go beyond Knabe's work \cite{K} on the two-dimensional AKLT model in terms of methods as well, and do not just rely on the improved computing power since 1988.)

\begin{figure}[t]
\begin{center}
\begin{tikzpicture}
\draw
(0:1)--(60:1)--(120:1)--(180:1)--(-120:1)--(-60:1)--cycle;
\foreach \j in {0, 60, 120, 180, -120, -60}{
\begin{scope}[shift=(\j:1)]
\draw (0:0)--(\j:1);
\draw[fill=black] (\j:1) circle (2.5pt);
\end{scope}}
\foreach \n in {0, 60, 120, 180, -120, -60}{
\draw[fill=red] (\n:1) circle (2.5pt);}
\end{tikzpicture}
\end{center}
\caption{The ``hexagonal sun'' subsystem $S$. The interactions along interior edges $\curly{E}_i$ (i.e., between the red vertices) are weighted by the parameter $a\geq 1$.}
\label{fig:sun}
\end{figure}
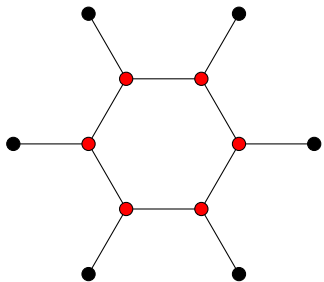

Let $S$ denote the hexagonal sun shown in Figure \ref{fig:sun}. We write $\curly{E}_i$ for the set of inner edges along the hexagon and $\curly{E}_o$ for the set of outer edges, the sun's rays. Let $a\geq 1$ be a parameter. Define the subsystem Hamiltonian
$$
H^S(a):=\sum_{e\in\curly{E}_i} ah_e+\sum_{e\in\curly{E}_o} h_e.
$$
Note that $H^S(a)$ is frustration-free and write $\gam^S(a)$ for its spectral gap. A routine modification of the proof of finite-size criteria then gives the following lower bound on the spectral gap of the two-dimensional model on any sufficiently nice subset $\Lam$ with periodic boundary conditions: $\gam^S(a)-\frac{a^2-2a+2}{2a+2}$, all mutiplied by a positive (and irrelevant) constant which only depends on $a\geq 1$.

For $a=1.4$, the gap threshold $\frac{a^2-2a+2}{2a+2}<0.242$. On the other hand, a direct computation shows that $\gam^S(1.4)>0.207$. Therefore, the weighted finite-size criterion fails, but rather narrowly: the relative difference between the subsystem gap and its threshold is less than $17\%$. 

The hexagonal sun is a subsystem of small size for which the method proposed here gets close to verifying the finite-size criterion and hence establishing the AKLT conjecture. We thus believe that larger cousins of the hexagonal sun are particularly suitable candidates for actually verifying a finite-size criterion in the full two-dimensional model. These larger systems are currently not amenable to what one could consider an essentially exact numerical calculation of the spectral gap via the Lanczos method. Still, we aim to explore an alternative numerical approach to these systems via DMRG in a future work.

\section*{Acknowledgments}
ML thanks Bruno Nachtergaele for encouragement and advice. AWS was supported by the NSF under grant DMR-1710170 and by a Simons Investigator award. The numerical calculations were carried out at Boston University's Shared Computing Cluster.

\end{document}